\documentclass[preprint,cite]{epl}
\usepackage{graphicx}
\usepackage{epsfig}
\usepackage[dvips]{color}

\newcommand{\red}[1]{\textcolor{black}{#1}}

\title{Depletion potentials near geometrically structured substrates}
\shorttitle{Depletion potentials near structured substrates}
\shortauthor{P. Bryk \etal}
\author{P. Bryk\inst{1,2} \and R. Roth\inst{1,2} \and M. Schoen\inst{3} \and S. 
Dietrich\inst{1,2}}
\institute{\inst{1} Max-Planck-Institut f{\"u}r Metallforschung --
Heisenbergstr. 3, D-70569 Stuttgart, Germany\\
\inst{2} Institut f{\"u}r Theoretische und Angewandte Physik, Universit{\"a}t 
Stuttgart --
Pfaffenwald\-ring 57, D-70569 Stuttgart, Germany\\
\inst{3} Stranski-Laboratorium f{\"u}r Physikalische und Theoretische Chemie, 
Sekretariat TC7,
Technische Universit{\"a}t Berlin -- Strasse des 17.Juni 124, D-10623 Berlin, 
Germany
}
\pacs{61.20.Gy}{Theory and models of liquid structure}
\pacs{82.70.Dd}{Colloids}

\begin{document}
\maketitle

\begin{abstract}
Using the recently developed so-called White Bear version of Rosenfeld's Fundamental
Measure Theory we calculate the depletion potentials between a hard-sphere
colloidal particle in a solvent of small hard spheres and simple models of
geometrically structured substrates: a right-angled wedge or edge. In the
wedge geometry, there is a strong attraction beyond the corresponding one
near a planar wall that significantly influences the structure of colloidal
suspensions in wedges. In accordance with an experimental study, for the edge
geometry we find  a free energy barrier of the order of several
$k_BT$ which repels a big colloidal particle from the edge.

\end{abstract}
When a colloidal suspension contains particles of different size,
depletion forces arise between big particles due to excluded volume effects.
If a colloidal suspension is prepared such that electrostatic
interactions are screened and only hard-core repulsive interactions are
left, these forces are attractive at small distances \cite{Asakura54} and can 
exhibit a potential
barrier and an oscillatory tail for large separations 
\cite{Dickman97,Biben96,Roth00a,Goetzelmann99}.
The same forces also arise if a big colloidal particle $b$ (diameter $\sigma_{b}$) 
suspended in a
solvent of small particles $s$ (diameter $\sigma_{s}$) approaches a wall.
The depletion forces for hard-sphere models of bulk colloidal
suspensions, as well as colloidal suspensions close to structureless walls have
been the subject of many theoretical studies.
Technical developments during the last decade made it possible to prepare 
well-defined,
monodisperse colloidal suspensions and with the help of video-microscopy
\cite{Crocker99} and total internal reflection microscopy 
\cite{Rudhardt98,Bechinger99,Helden03}
such depletion potentials can be now measured directly and compared quantitatively
with theoretical predictions.

On the other hand, the behavior of colloidal suspensions
close to {\it structured} walls and resulting depletion potentials have
not yet been investigated in depth.
Such situations arise naturally 
for example close to a corner in a container for colloidal suspensions.
Geometrically structured substrates with a dedicated
design have been used to control the growth of colloidal crystals \cite{Yin02}.
In this context Dinsmore et al. \cite{Dinsmore96} studied the motion of colloidal
"hard-spheres`` near the edge of a terrace by means of video-microscopy.
In the direction perpendicular to the edge they
found a free-energy barrier approximately two times the mean thermal energy.
Dinsmore et al. suggested that localized
entropic force-fields created by various structures of the substrate can be used
to control the movement of colloidal particles and, consequently, could be
used to create highly ordered arrays of colloidal particles.

Recently, Kinoshita et al. \cite{Kinoshita02a,Kinoshita02b} have
studied depletion potentials of hard spheres close to
the substrate exhibiting various geometrical features by means of
first-order Ornstein-Zernike (FOOZ) integral equations. Using correlation
functions of the bulk hard-sphere fluid as input,
they calculated the potential of mean force between a big particle and
an edge or hemispherical hole in a planar substrate (mimicking key and lock steric 
interactions
between macromolecules). However, FOOZ theory assumes that the
correlation functions close to the surface are the same as in the
bulk, which gives rise to discrepancies in predicting packing effects as compared 
to simulation
data. For geometrically structured surfaces, where these packing effects may be 
very
pronounced like, e.g. in a wedge,  the FOOZ approach is expected to be even less 
reliable.

Depletion forces in the bulk or close to planar walls have been
studied also by means of molecular dynamics \cite{Dickman97}
and Monte Carlo (MC) simulations \cite{Biben96}. While being computationally quite 
demanding,
these studies serve as a useful test for various theories.
However, the study of the fluid structure near geometrically structured substrates 
is
computationally even more costly than for planar walls so that there are only few 
such studies
\cite{Diestler00,Schoen02,Schoen97}.

A versatile approach to study depletion potentials close to arbitrarily
shaped substrates has been proposed in Refs. \cite{Roth00a} and 
\cite{Goetzelmann99}.
According to this theory the depletion potential is evaluated using the
potential distribution theorem \cite{Roth00a,Goetzelmann99,Henderson83,Widom63}
on the basis of the density profile of the small spheres close to the substrate 
without
the big sphere. This method is valid for all geometries of the substrate, but 
explicit calculations so far
have been carried out only for structureless planar walls \cite{Roth00a} or
spherically curved substrates \cite{Roth99}.

Here we apply the aforementioned scheme to
calculate the depletion potential of a single big hard sphere close to a 
right-angled
wedge or edge. Within this approach it is important to determine
accurately the density profile $\rho_{s}({\bf r})$ of the small spheres close to
the edge or wedge where it varies most rapidly.

The calculation of the density profile of a hard-sphere fluid close to planar or 
spherical
structureless walls constitutes a standard problem in the
statistical mechanics of fluids which lends itself to be solved by density 
functional theory (DFT).
It has emerged that \red{Rosenfeld's} Fundamental Measure Theory (FMT) 
\cite{Rosenfeld89}
is one of the most accurate DFTs if compared with simulations for either
one-component, binary \cite{Roth00b} or polydisperse \cite{Pizio01}
hard-sphere fluids close to a hard-wall. However, for complex geometries like
wedges or edges there are only a few comparisons of DFT and simulations
available \cite{Henderson98,Jagannathan02}. Thus before calculating the
depletion potentials we test the accuracy of the DFT
against MC simulations.

\begin{figure}
\centering\includegraphics[width=6cm,clip]{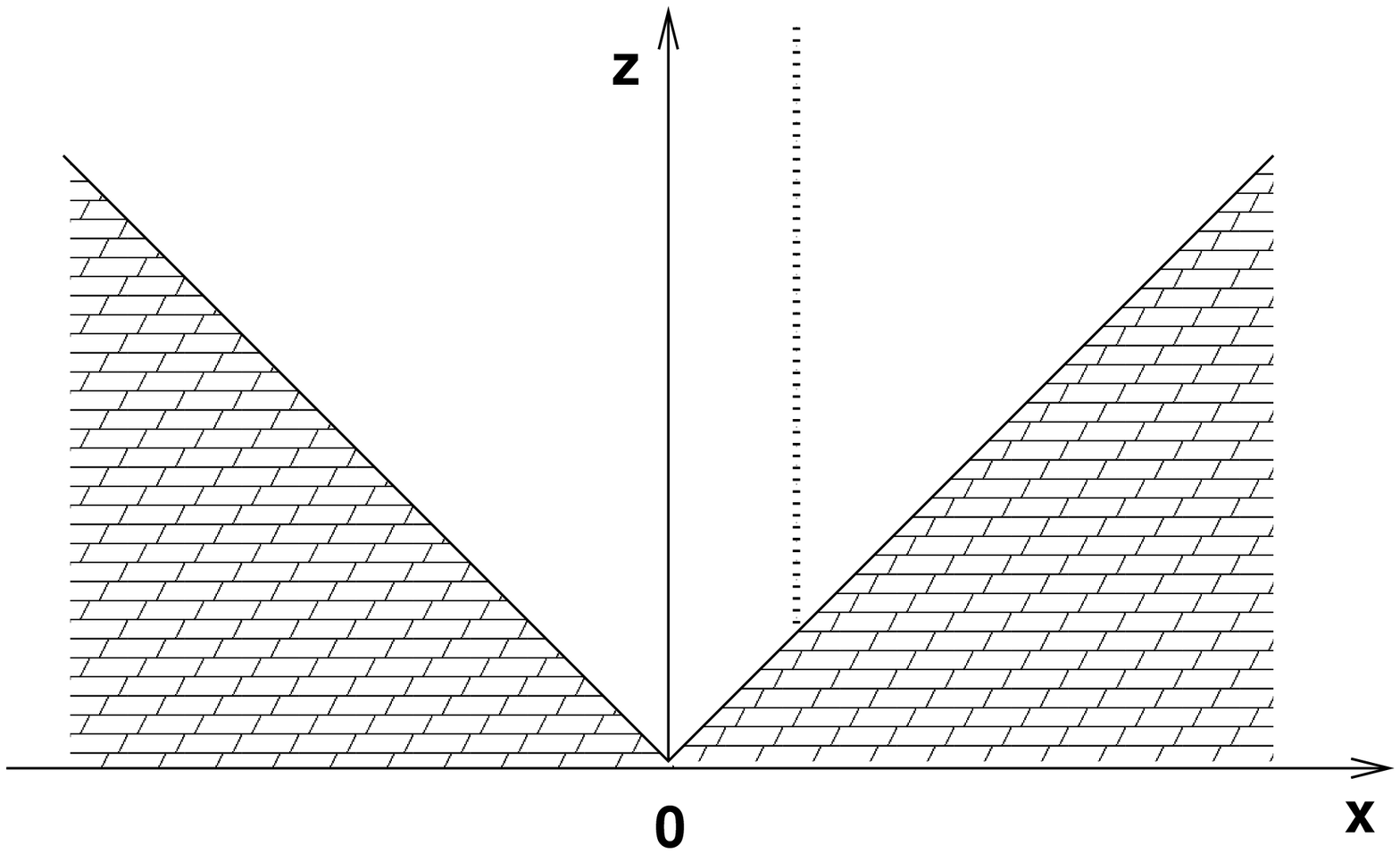}
\includegraphics[width=6cm,clip]{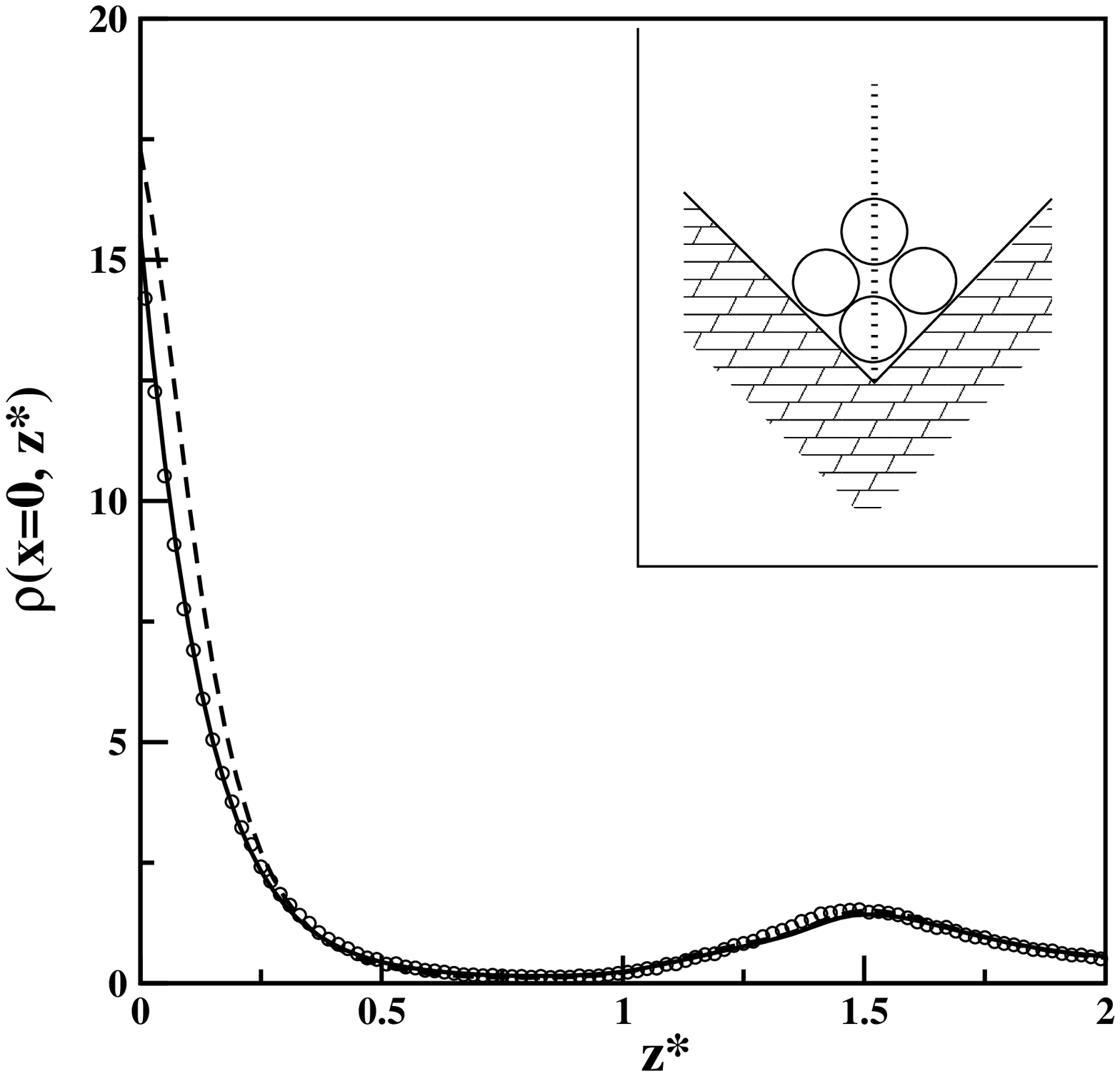}
\caption{\label{fig:1} Schematic representation of a right-angled wedge 
$z_{w}(x)=|x|$.
The dotted line indicates a possible cut along which the density profiles are 
presented.}
\caption{\label{fig:2} Density profile $\rho(x=0,z^{\ast})$ from MC (circles),
$RF$ version (dashed line) and $WB$ version of
FMT (solid line) for a bulk density $\rho_s=0.7$, with 
$z^{\ast}=\frac{(z-z_{w})}{\sigma_{s}}-\frac{1}{\sqrt{2}}$.
The inset illustrates this density distribution by showing the corresponding 
typical
particle configuration. Here and below lengths and number densities are given in 
units
of $\sigma_{s}$ and $\sigma_{s}^{-3}$, respectively.}
\end{figure}

The key approximation in every density functional theory resides in an expression 
for
the excess (over the ideal gas) intrinsic
free energy  ${\cal F}_{ex}^{(hs)}[\rho_{i}({\bf r})]$ of an inhomogeneous fluid 
as
a functional of local densities $\rho_{i}({\bf r})$.
Within the framework of FMT \cite{Rosenfeld89} one has
\begin{equation}
\beta{\cal F}_{ex}^{(hs)}[\{\rho_{i}\}]=\int d^{3}r \;\Phi(\{n_{\alpha}\}) \;,
\end{equation}
where $\beta=(k_BT)^{-1}$, $k_B$ is the Boltzmann constant, and $T$ is the 
temperature.
The excess free energy density $\Phi$ is a function of weighted densities 
\begin{equation}
n_{\alpha}({\bf r})=\sum_{i=s,b}\int d^{3}r' \rho_{i}({\bf r}')~
w^{(\alpha)}_{i}({\bf r}-{\bf r}') \;,
\end{equation}
with six different geometrical weight functions $w^{(\alpha)}_{i}$ per component 
$i$
(four scalar and two vector-like).
In the present problem all 
quantities are translationally invariant
along the wedge or edge which we choose as the $y$-direction.
In order to calculate weighted densities $n_{\alpha}(x,z)$ in an efficient way
we have applied the two-dimensional Fast Fourier Transform.

From a variety of expressions for $\Phi$ here we select two of them.
The first one is the original Rosenfeld functional ($RF$) \cite{Rosenfeld89}
\begin{equation}
\Phi^{(RF)}(\{n_{\alpha}\})=-n_{0} \ln (1-n_{3})+\frac{n_{1}n_{2}-{\bf n}_{1}\cdot
{\bf n}_{2}}{1-n_{3}} +\frac{n_{2}^{3}-3n_{2}{\bf n}_{2}\cdot{\bf n}_{2}}
{24\pi (1-n_{3})^{2}} \;,
\end{equation}
based on the Percus-Yevick compressibility equation of state. As pointed out in 
Ref.~\cite{Rosenfeld97}
this approximation does not render a freezing transition.
Our second choice is based on  the Boublik-Mansoori-Carnahan-Starling-Leland 
equation of state
\cite{Boublik70,Mansoori71} as proposed by Tarazona \cite{Tarazona02} and Roth 
\etal \cite{Roth02}
(see also Ref.~\cite{Yu02} where the 
same functional has been proposed, too). Moreover this choice implements a modification proposed in 
Ref.~\cite{Rosenfeld97}
(the so-called antisymmetrized form, $Q=3$) that allows for the occurrence of a
freezing transition:
\begin{equation}
\Phi^{(WB)}(\{n_{\alpha}\})=-n_{0} \ln (1-n_{3})+\frac{n_{1}n_{2}-{\bf n}_{1}\cdot
{\bf n}_{2}}{1-n_{3}} +n_{2}^{3}(1-\xi^2)^3\frac{n_{3}+(1-n_{3})^{2}\ln (1-n_{3})}
{36\pi n_{3}^{2}(1-n_{3})^{2}} \;,
\end{equation}
where $\xi({\bf r})=|{\bf n}_{2}({\bf r})|/n_2({\bf r})$.
Following the nomenclature in Ref.~\cite{Roth02} this DFT version is denoted
$WB$ ("White Bear``). The density profiles are obtained by solving the
Euler-Lagrange equations corresponding to the minimum of the grand potential 
functional.

The Monte Carlo simulations have been performed in a grand canonical ensemble
according to a procedure described in detail in Ref.~\cite{Schoen97}. However,
we emphasize that unlike in Ref.~\cite{Schoen97} the MC-generated
density profiles presented below have not been smoothed but are instead the
''raw`` data generated in MC simulations with no further post-simulation
processing applied.

The comparison of the DFT and MC density profiles has been carried out only for 
the wedge geometry,
because in this geometry packing effects are much more pronounced than for the
edge and thus presents the more severe challenge for DFT.
The dependence of the density profile $\rho_{s}=\rho_{s}(x,z)$ on two spatial
variables is illustrated by vertical cuts along the lines $x=\mbox{const}$ (see 
fig.~\ref{fig:1}).
In the following all length scales are expressed in units of $\sigma_s$, the 
diameter
of the small spheres, and the number densities in units of $\sigma_{s}^{-3}$.

\begin{figure}
\centering\includegraphics[width=6.5cm,clip]{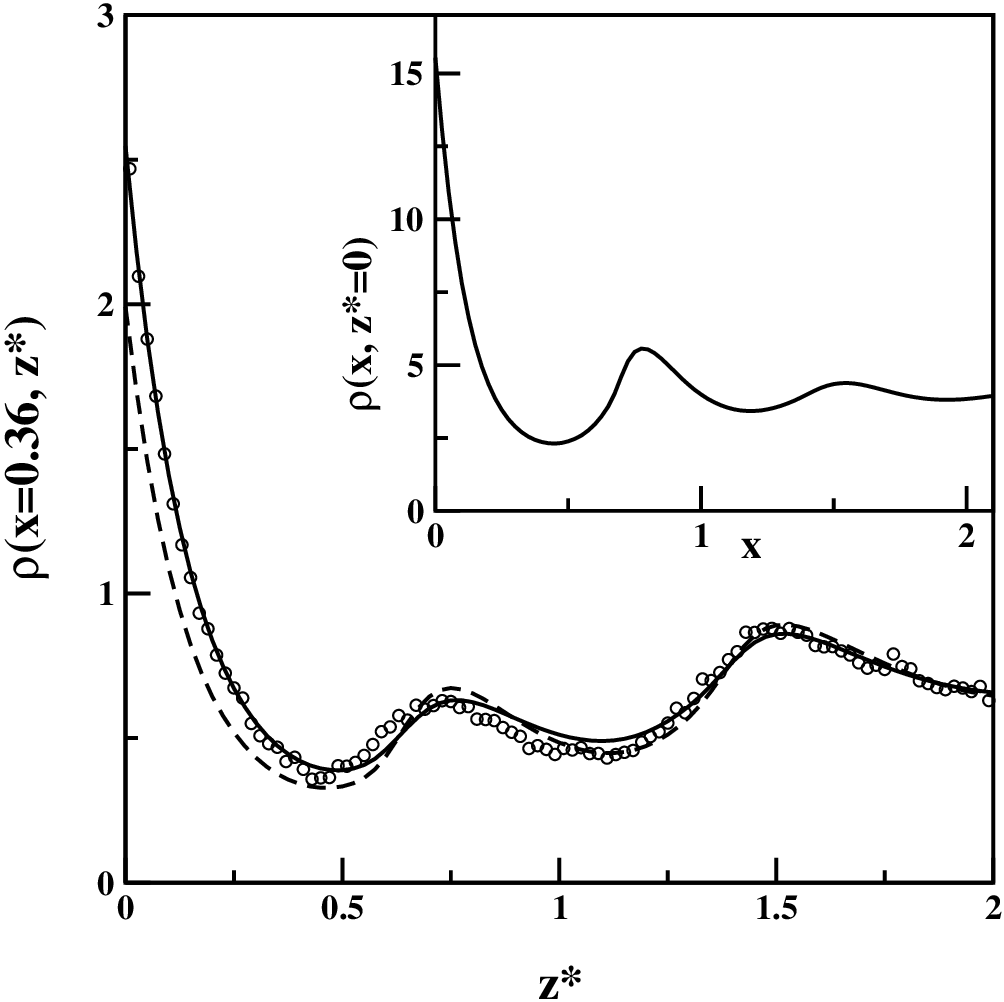}
\includegraphics[width=6.5cm,clip]{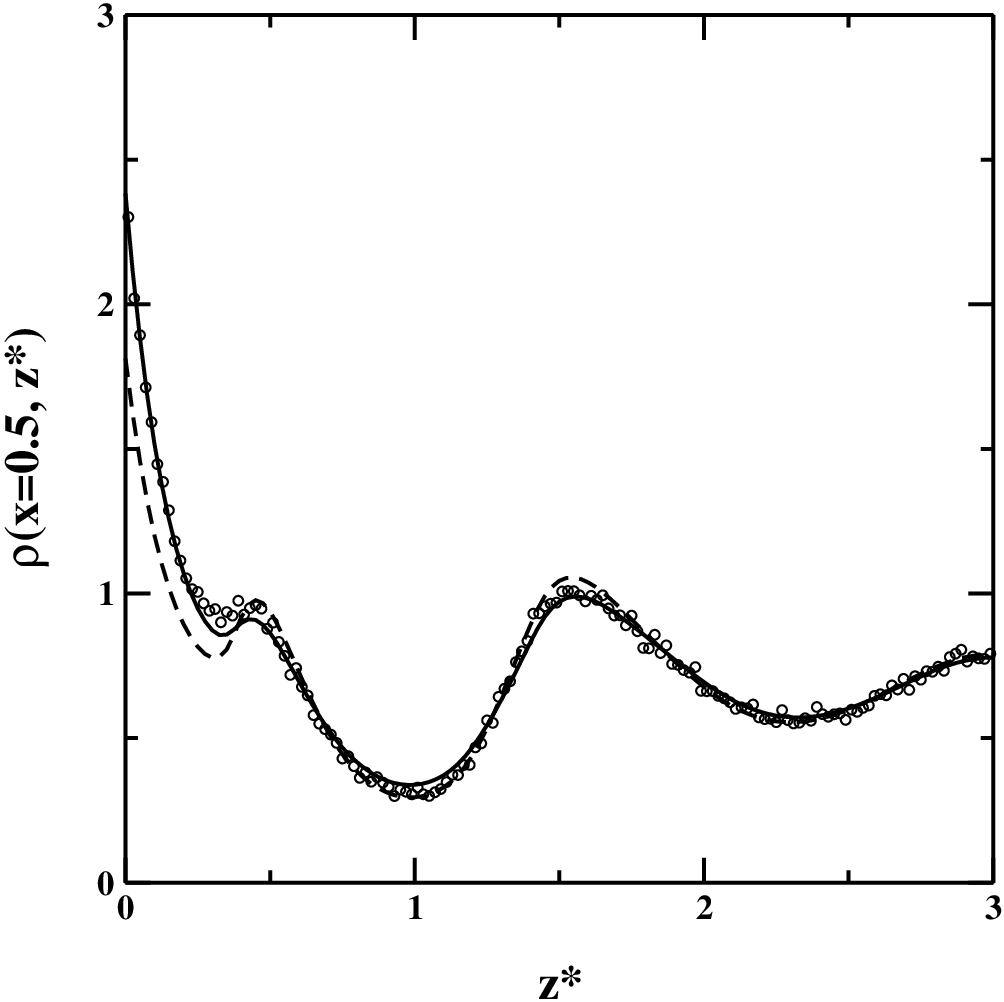}
\caption{\label{fig:3} The density profile $\rho(x=0.36,z^{\ast})$
from MC simulations (circles), $RF$ version (dashed line), and
$WB$ version (solid lines) evaluated for the bulk density
$\rho_s=0.7$. The inset shows the contact value of the density profile resulting
from the White Bear version as function of $x$.}
\caption{\label{fig:4} The density profile $\rho(x=0.5,z^{\ast})$ from MC 
simulations
(circles), $RF$ version (dashed line), and $WB$ version (solid line) evaluated
for the bulk density $\rho_s=0.7$.}
\end{figure}

Figures \ref{fig:2}-\ref{fig:4} show such MC density profiles (circles) and
the corresponding ones from DFT (dashed and solid lines).
The comparison with the MC data shows that the $WB$ version represents a
noticeable improvement over the $RF$ version in particular for
$z^{\ast}=\frac{z-z_{w}}{\sigma_{s}}-\frac{1}{\sqrt 2}<0.5$. Along the line $x=0$ (see fig.~\ref{fig:2})
the contact value of the profile resulting from the $WB$ version is 15.54 while
from $RF$ 17.29, i.e., a difference of ca. 10\%. Also for $x=0.36$ (see 
fig.~\ref{fig:3}) the
$WB$ version agrees better with the MC data than $RF$ again in particular for 
$z^{\ast}<0.5$.
The contact value $\rho_{s}^{(WB)}(x=0.36,z^{\ast}=0)=2.55$ differs from
$\rho_{s}^{(RF)}(x=0.36,z^{\ast}=0)=1.99$ by ca. 20\%.

The large variations (between 15.5 and 2.5) of the contact
densities for $0\le x\le0.5$ (see the inset in fig.~\ref{fig:3}) illustrate
the strong packing effects in this system. In fig.~\ref{fig:4} for $x=0.5$
a small plateau develops around $z^{\ast}=0.5$. The $RF$ version exaggerates
this feature considerably and leads to a contact value $\rho_{s}^{(RF)}
(x=0.5,z^{\ast}=0)=1.81$ whereas the $WB$ version reproduces the plateau more 
accurately
leading to a contact value $\rho_{s}^{(WB)}(x=0.5,z^{\ast}=0)=2.38$.

We refrain from comparing directly contact values from DFT with those from
MC because in the MC simulations $\rho\left(x,z\right)$ has to be calculated as a
histogram on a two-dimensional grid. Even though the mesh width
$\delta x=\delta z=0.02$ of this grid is small, an extrapolation scheme is 
required to obtain an
estimate for the contact value. This extrapolation scheme is particularly delicate
near the core of the wedge where $\rho\left(x,z\right)$ varies
rapidly. However, inspection of figs.~\ref{fig:2}-\ref{fig:4}
leads to the expectation that the $WB$ version of DFT reproduces the actual
contact values rather accurately. We have carried out calculations also for lower
bulk densities finding a similar (or even better) agreement
between DFT and simulations. Thus we conclude that FMT
predicts accurately the structure  of the hard-sphere fluid in a hard
wedge; its $WB$ version improves the agreement with simulations near the
core of the wedge, where packing effects are particularly pronounced.

Reassured by the quality of the density profiles resulting from DFT, we
have calculated the depletion potentials $W$ near a hard wedge and edge
based on the $WB$ version of DFT.
We use the fact that in the limit of a dilute solution of big particles $\beta W$ can be
calculated from the density distribution of the small spheres unperturbed by
the presence of the big one. It has been shown \cite{Roth00a,Goetzelmann99}
that in the dilute limit of the big spheres the depletion potential of species $b$ 
at a
point ${\bf r}$  can be expressed in terms of the difference between
the one-particle direct correlation
function \red{$c_{b}^{(1)}$} in the bulk (${\bf r}\to\infty$) and at ${\bf r}$:
\begin{equation}
\beta W({\bf r})=\lim_{\rho_{b}\to 0}
\left(c_{b}^{(1)}({\bf r}\to\infty)- c_{b}^{(1)}({\bf r})\right)\:
\end{equation}
with $c_b^{(1)}({\bf r})=-\delta {\cal F}_{ex}/ \delta\rho_b({
\bf r})$ within DFT \cite{Evans79}.

\begin{figure}
\begin{center}
\includegraphics[width=6.5cm,clip]{f_d07_edge_new.eps}
\includegraphics[width=6.5cm,clip]{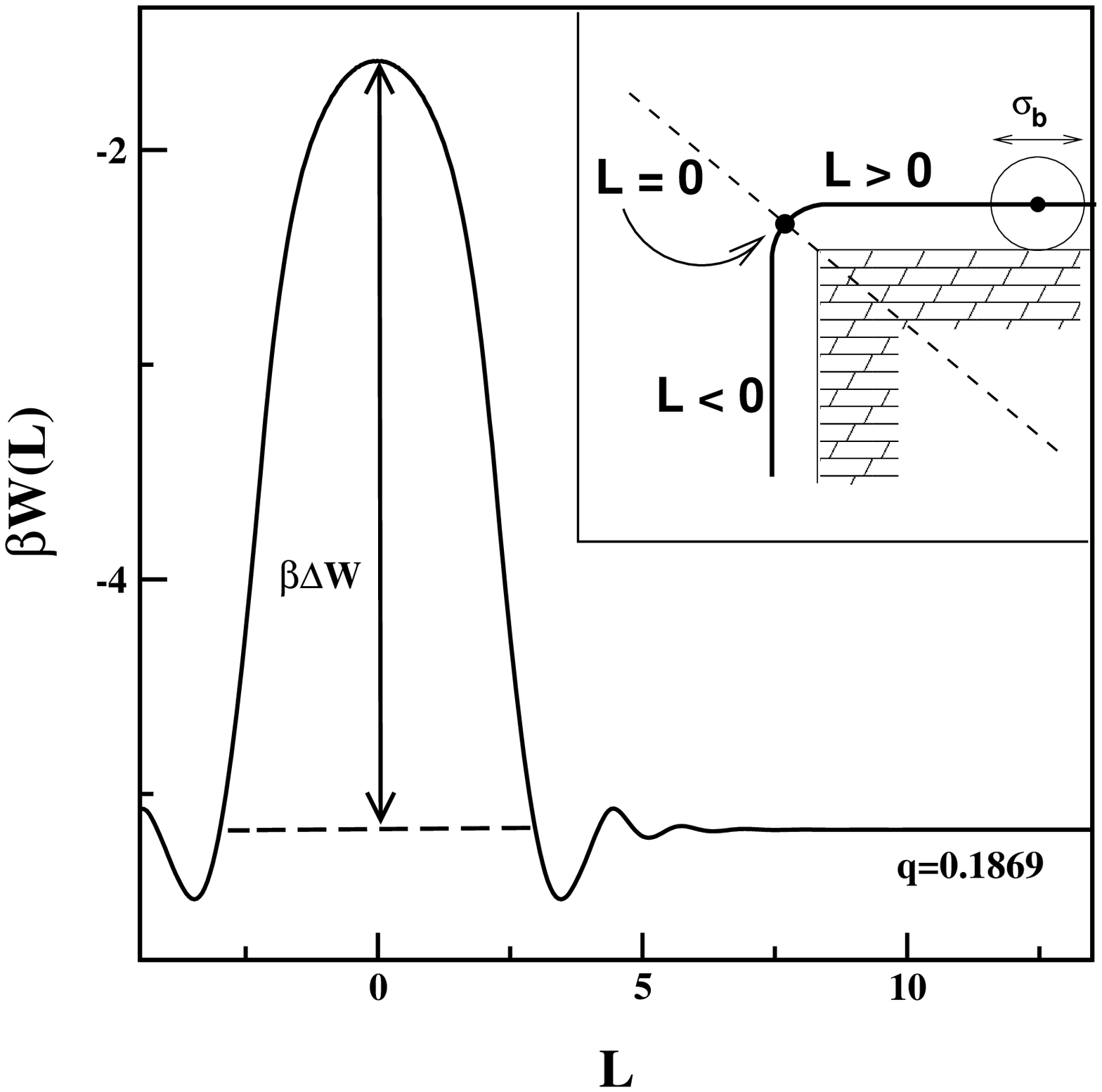}
\end{center}
\caption{\label{fig:5} The depletion potential of the big colloidal particle
close to the edge along the line of closest contact (see the inset in 
fig.(\ref{fig:6})).
The packing fraction of the small spheres is $\eta=0.366$. The dashed, solid, and
dashed-dotted lines correspond to the size ratios
$q\equiv\sigma_{s}/\sigma_{b}=0.333$, 0.2, and 0.1, respectively.}
\caption{\label{fig:6} The depletion potential along the line of closest contact
between the big colloidal particle and the 90$^{\circ}$ right-angled edge (inset) 
evaluated for
$\sigma_{b}=5.35\sigma_{s}$ and $\eta=0.3$}
\end{figure}

In fig.~\ref{fig:5} we show $\beta W$ for size ratios $q^{-1}\equiv\sigma_b/
\sigma_s=3$, 5, and 10 at a hard edge for a bulk packing fraction of
the small spheres $\eta=\frac{\pi}{6}\rho_{s}\sigma^{3}_{s}=0.366$.
The depletion potential is evaluated along the line of contact between
the big sphere and the edge (see the inset in fig.~\ref{fig:6});
$L$ parametrizes a path on this line.
For $L\to\pm\infty$ the depletion potential between a planar
hard wall and a big sphere is recovered. Upon approaching
the edge, oscillations in the contact value of $\beta W$ occur.
At the corner there is a pronounced maximum in the (still attractive) depletion 
potential.
The value of $\beta W$ at the maximum, i.e., at $L=0$ equals
$-1.5$, $-1.6$ and $-2.0$ for $q=0.333$, 0.2, and 0.1, respectively.  This maximum
represents an effective repulsive barrier repelling a big colloidal particle
approaching the edge from the side and practically preventing it from passing 
around the corner.
The difference $\beta\Delta W$ between the depletion potential at the planar hard 
wall and at the
corner is a quantitative measure of the barrier height.
The barrier height increases upon
decreasing $q$ and for the system shown in fig~\ref{fig:5} it equals 2.1, 3.8, and 
8.0, respectively.

\begin{figure}
\centering\includegraphics[width=6.5cm,clip]{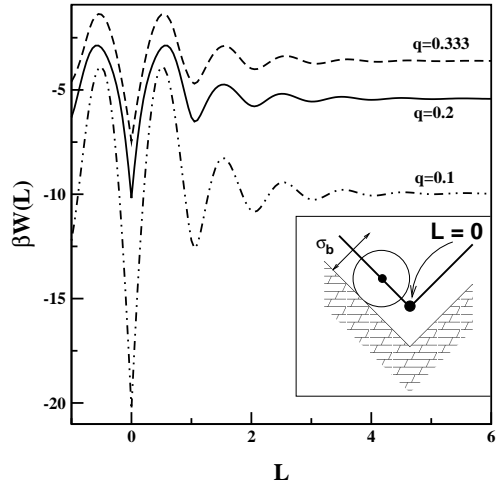}
\caption{\label{fig:7} The depletion potential of the big colloidal particle
close to the wedge taken along the line of closest contact.
The packing fraction of the small spheres is $\eta=0.366$. The dashed, solid and 
dotted-dashed lines
correspond to the size ratios $q=0.333$, 0.5 and 0.1, respectively.}
\end{figure}

In fig.~\ref{fig:6} we present $\beta W$ along the line of contact for a system
mimicking the experimental system studied by
Dinsmore et al. \cite{Dinsmore96}. Theoretically we find $\beta\Delta
W\simeq3.6$ which is larger than the corresponding result obtained from
the Asakura-Oosawa approximation ($\beta\Delta W\simeq3$, \cite{Dinsmore96})
which treats the solvent as an ideal gas.
The experimental value quoted for the barrier height $(\simeq 2)$ 
\cite{Dinsmore96} is
smaller than both theoretical results. This is expected because for the 
theoretical
value the big particle is assumed to move along the line of contact with 
the wall
whereas experimentally on average some of the big particles monitored 
experimentally happen to be located slightly above the surface.
$\beta\Delta W$ reaches a value of 2 at a normal distance of about
0.2$\sigma_s$.
The depletion potentials shown in figs.~\ref{fig:5} and \ref{fig:6}
exhibit an oscillatory behavior due to correlation effects missed by the 
Asakura-Oosawa
approximation which holds only in the limit $\eta\to 0$. These correlations lead
to a minimum of $\beta W$ at $L\simeq3.5$ next to the main peak (see 
fig.~\ref{fig:6}),
which is the global minimum and is also visible in the experimental depletion potential
\cite{Dinsmore96}.

In fig.~\ref{fig:7} we show $\beta W$ along the line of contact for a single big 
hard sphere in a
right-angled hard wedge for $\eta=0.366$ and $q=0.333$, 0.2, and 0.1. As in
fig.~\ref{fig:5} we focus on the depletion potential along the line of contact 
between
the big sphere and the wedge surface (see the inset in fig.~\ref{fig:7}).
Again in the limit $L\to\pm\infty$ the contact value of $\beta W$ between a
planar hard wall and a big hard sphere is recovered. Near the core of the wedge
oscillations occur which are more pronounced than
for the corresponding edge cases. Attraction is strongest directly in the corner
where $\beta W=-7.5$, $-10.2$, and $-20.3$ for $q=0.333$, 0.2, and 0.1, 
respectively.
This is roughly twice the corresponding contact values at a planar hard wall
(-3.6, -5.4, and -10.0). This potential is expected to promote highly ordered 
structures of big colloids
in the core of the wedge. The pronounced oscillations of the depletion potential
upon approaching the core of the wedge along the walls can lead to 
additional ordering.

We have studied depletion potentials near geometrically structured substrates.
It has turned out that the recently developed ``White Bear'' version \cite{Roth02}
of Fundamental Measure Theory in its antisymmetrized form \cite{Rosenfeld97}
is able to reproduce accurately the structure of the hard-sphere fluid in a
hard wedge as obtained from Monte Carlo simulations.
wall.
The free energy barrier repelling a big colloidal particle from a substrate edge 
can be
of the order of several $k_B T$ in accordance with experiments \cite{Dinsmore96}.
In conclusion, the results presented
in this letter demonstrate that geometrically structured substrates
create strong and oscillatory entropic force fields that may significantly affect
the structure and hence the properties of colloidal suspensions near their 
container walls.


\begin{thebibliography}{0}
\bibitem{Asakura54}
  \Name{Asakura S. \and Oosawa F.}
  \REVIEW{J. Chem. Phys.}{22}{1954}{1255}.
\bibitem{Dickman97}
  \Name{Dickman R., Attard P. \and Simonian V.}
  \REVIEW{J. Chem. Phys.}{107}{1997}{205}.
\bibitem{Biben96}
  \Name{Biben T., Bladon P.  \and Frenkel D.}
  \REVIEW{J. Phys. Condens. Matter}{8}{1996}{10799}.
\bibitem{Roth00a}
  \Name{Roth R., Evans R. \and  Dietrich S.}
  \REVIEW{Phys. Rev. E}{62}{2000}{5360}.
\bibitem{Goetzelmann99}
\Name{G\"{o}tzelmann B., Roth R., Dietrich S., Dijkstra M. \and Evans R.}
\REVIEW{Europhys. Lett.}{47}{1999}{398}.

\bibitem{Crocker99}
  \Name{Crocker J. C., Matteo J. A., Dinsmore A. D. \and  Yodh A. G.}
  \REVIEW{Phys. Rev. Lett.}{82}{1999}{4352}.
\bibitem{Rudhardt98}
  \Name{Rudhardt D., Bechinger C. \and Leiderer P.}
  \REVIEW{Phys. Rev. Lett.}{81}{1998}{1330}.
\bibitem{Bechinger99}
  \Name{Bechinger C., Rudhardt D., Leiderer P., Roth R. \and Dietrich S.}
  \REVIEW{Phys. Rev. Lett.}{83}{1999}{3960}.
\bibitem{Helden03}
  \Name{Helden L., Roth R., Koenderink G. H., Leiderer P. \and Bechinger C.}
  \REVIEW{Phys. Rev. Lett.}{90}{48301}{2003}.
\bibitem{Yin02}
  \Name{Yin Y. \and Xia Y.}
  \REVIEW{Adv. Mater.}{14}{2002}{605}.
\bibitem{Dinsmore96}
  \Name{Dinsmore A. D., Yodh A. G. \and Pine D. J.}
  \REVIEW{Nature}{383}{1996}{239}.
\bibitem{Kinoshita02a}
  \Name{Kinoshita M. \and Oguni T.}
  \REVIEW{Chem. Phys. Lett.}{351}{2002}{79}.
\bibitem{Kinoshita02b}
  \Name{Kinoshita M.}
  \REVIEW{J. Chem. Phys.}{116}{2002}{3493}.
\bibitem{Diestler00}
  \Name{Diestler D. J. \and Schoen M.}
  \REVIEW{Phys. Rev. E}{62}{2000}{6615}.
\bibitem{Schoen02}
  \Name{Schoen M.}
  \REVIEW{Colloids and Surf. A: Physicochem. Eng. Asp.}{206}{2002}{253}.
\bibitem{Schoen97}
  \Name{Schoen M. \and Dietrich S.}
  \REVIEW{Phys. Rev. E}{56}{1997}{499}.
\bibitem{Henderson83}
  \Name{Henderson J. R.}
  \REVIEW{Mol. Phys.}{50}{1983}{741}.
\bibitem{Widom63}
  \Name{Widom B.}
  \REVIEW{J. Chem. Phys.}{39}{1963}{2808}.
\bibitem{Roth99}
  \Name{Roth R., G\"{o}tzelmann B. \and  Dietrich S.}
  \REVIEW{Phys. Rev. Lett.}{83}{1999}{448}.
\bibitem{Rosenfeld89}
  \Name{Rosenfeld Y.}
  \REVIEW{Phys. Rev. Lett.}{63}{1989}{980}.
\bibitem{Roth00b}
  \Name{Roth R. \and  Dietrich S.}
  \REVIEW{Phys. Rev. E}{62}{2000}{6926}.
\bibitem{Pizio01}
  \Name{Pizio O., Patrykiejew A. \and Soko{\l}owski S.}
  \REVIEW{Mol. Phys.}{99}{2001}{57}.
\bibitem{Henderson98}
  \Name{Henderson D., Soko{\l}owski S. \and Wasan D. T.}
  \REVIEW{Phys. Rev. E}{57}{1998}{5539}.
\bibitem{Jagannathan02}
  \Name{Jagannathan K. \and Yethiraj A.}
  \REVIEW{J. Chem. Phys.}{116}{2002}{5795}.
\bibitem{Rosenfeld97}
  \Name{Rosenfeld Y., Schmidt M., L\"owen H. \and Tarazona P.}
  \REVIEW{Phys. Rev. E}{55}{1997}{4245}.
\bibitem{Boublik70}
  \Name{Boublik T.}
  \REVIEW{J. Chem. Phys.}{53}{1970}{471}.
\bibitem{Mansoori71}
  \Name{Mansoori G. A., Carnahan N. F., Starling K. E. \and Leland T. W. Jr.}
  \REVIEW{J. Chem. Phys.}{54}{1971}{1523}.
\bibitem{Tarazona02}
  \Name{Tarazona P.}
  \REVIEW{Physica A}{306}{2002}{243}.
\bibitem{Roth02}
  \Name{Roth R., Evans R., Lang. A. \and  Kahl. G.}
  \REVIEW{J. Phys. Condens Matter}{14}{2002}{12063}.
\bibitem{Yu02}
  \Name{Yu Y.-X. \and  Wu J.}
  \REVIEW{J. Chem. Phys.}{117}{2002}{10165}.

\bibitem{Evans79}
  \Name{Evans R.}
  \REVIEW{Adv. Phys.}{28}{1979}{143}.
\end{thebibliography}
\end{document}